\documentclass
[superscriptaddress,amsmath,amssymb,aps,pra,lengthcheck]{revtex4-1}%
\usepackage{graphicx}
\usepackage{dcolumn}
\usepackage{bm}
\usepackage{hyperref}
\usepackage{amsmath}
\usepackage{amssymb}
\usepackage{dsfont}%
\usepackage{amsfonts}
\newcommand{\abs}[1]{\left| #1 \right|}

\def\vec#1{{\boldsymbol{#1}}}

\begin{document}

\title{Robustness of Fractional Quantum Hall States with Dipolar Atoms in
Artificial Gauge Fields}

\author{T. Gra\ss }
\affiliation{ICFO-Institut de Ci\`encies
Fot\`oniques, Mediterranean
Technology Park, 08860 Castelldefels (Barcelona), Spain}
\author{M. A. Baranov}
\affiliation{Institute for Theoretical Physics, University of Innsbruck, A-6020
Innsbruck, Austria}
\affiliation{Institute for Quantum Optics and Quantum Information of the
Austrian Academy of Sciences, A-6020 Innsbruck, Austria}
\affiliation{RRC ``Kurchatov Institute'', Kurchatov Square 1, 123182 Moscow, Russia}
\author{M. Lewenstein}
\affiliation{ICFO-Institut de Ci\`encies
Fot\`oniques, Mediterranean
Technology Park, 08860 Castelldefels (Barcelona), Spain}
\affiliation{ICREA-Instituci\'o Catalana de Recerca i Estudis Avan\c cats, Lluis Companys 23, 08010 Barcelona, Spain}

\begin{abstract}
The robustness of fractional quantum Hall states is measured as the energy gap
separating the Laughlin ground-state from excitations. Using thermodynamic
approximations for the correlation functions of the Laughlin state and the
quasihole state, we evaluate the gap in a two-dimensional system of dipolar
atoms exposed to an artificial gauge field. For Abelian fields, our results
agree well with the results of exact diagonalization for small systems, but
indicate that the large value of the gap predicted in [Phys. Rev. Lett.
\textbf{94}, 070404 (2005)] was overestimated. However, we are able to show
that the small gap found in the Abelian scenario is dramatically increased if
we turn to non-Abelian fields squeezing the Landau levels.
\end{abstract}

\date{\today}
\maketitle

\section{Introduction}

Thirty years after its discovery in GaAs-AlGaAs heterojunctions \cite{tsui},
the fractional quantum Hall effect (FQHE) still remains to be the topic of
current research, as nowadays this former solid-state phenomenon is
re-discovered within the scope of quantum gases in two dimensions. Proposals
to achieve the crucial ingredient, a perpendicular magnetic field, within such
systems of neutral particles are at hand: Conceptually the simplest of them is
a rotation of the system
\cite{wilkin,*schweikhard,*rezayi,*fetter-revmod,*cooper-aip,baranov}, where
the strength of the artificial magnetic field can be tuned by the frequency of
rotation. However, addressing the regime where only the lowest Landau level
(LLL) is occupied, and at the same time guaranteeing stability of the atomic
cloud requires a delicate balance between rotation and trap frequency, making
an experimental realization of this proposal extremely hard.

An alternative way which circumvents this problem is to generate artificial
fields by implementing a laser-assisted tunneling scheme within an optical
lattice \cite{jakschzoller, *mueller, *sorensen, *hafezi}. The idea is that
during a hopping process stimulated by a laser field, the atom picks up the
phase of the laser, which effectively simulates the action of a magnetic
vector potential. In a very similar way, a $U(1)$ Berry phase mimicking a
gauge field can be inscribed into continuous systems via atom-laser coupling
\cite{juzeli2004, *guenter, *dalibard-gerbier, *dalibard}. Furthermore, if the
coupling involves more than two atomic states, it is possible to create a
space-dependent degenerate subspace, which can be understood in terms of a
non-Abelian Berry phase communicating between the degenerate atomic states. In
this way, the atom-laser coupling scheme is generalized for synthesizing also
non-Abelian gauge fields \cite{fleischhauer}. The same can be achieved in
optical lattices, if the laser-assisted tunneling is made sensitive on an
additional, internal degree of freedom of the atoms \cite{osterloh}. These new
possibilities have directed the attention to FQHE in such non-Abelian
scenarios \cite{trombettoni}. The practical feasibility of using lasers to
implement artificial magnetic fields has been shown in pioneering experiments
\cite{lin, *spielmanPRL}. Furthermore, very recently the implementation of
spin-orbit coupling within a Bose-Einstein condensate can be considered an
$SU(2)$ gauge field realized in cold atoms \cite{spielman-sobec}.

The main motivation for seeking new realizations of the old fractional quantum
Hall effect is given by the intriguing quasiparticles which occur as the
excitations of fractional quantum Hall systems: Being neither bosons nor
fermions, these so-called anyons behave exotically under interchange
\cite{wilczek}, as instead of gaining simply a sign factor, a phase is
obtained, which in degenerate anyonic systems may even be an element of some
non-Abelian group. Due to this property, together with their topological and
thus robust nature, quantum Hall states are especially interesting for quantum
computation \cite{nayak}.

It is therefore most relevant to find fractional quantum Hall systems, where a
large energy gap separating the ground state from the excited states
guarantees a high degree of robustness.

Considering dipolar atoms in a rotating trap, Ref. \cite{baranov} claims to
have achieved this. In this Paper, we consider once again this scenario and
find a much smaller gap. This finding directs our attention to the question if
a robust FQHE might instead be realizable in a non-Abelian environment. From
Refs. \cite{anjaPRL,*anjaPRA} it is known that in non-interacting systems the
non-Abelian gauge field strongly influences the nature of the integer quantum
Hall effect. By choosing an anisotropic field configuration it becomes possible
to squeeze the Landau levels. Applying to such a squeezed scenario the
thermodynamic approximation \cite{girvin-g0, girvin} used in the Abelian case
in Ref. \cite{baranov}, we show in this Paper, that a dramatic increase of the
energy gap can be achieved.

As in dipolar systems it turns out that the gap crucially depend on how it is
defined, we first discuss this matter in Section \ref{II}, which we conclude
with a precise definition of the gap. Subsequently, we describe in Section
\ref{ii} an analytic approximation allowing to evaluate this gap in the
thermodynamic limit. Section \ref{III} is dedicated to specifying the gauge
potential. Here we show that Abelian and non-Abelian fields can be treated on
the same footing. Finally, the concrete calculation and its results are
described in Section \ref{IV}.

\section{\label{II} Fractional Quantum Hall Effect and definition of the gap}

The physics of non-interacting particles confined to a plane under the
influence of a perpendicular magnetic field is understood in terms of the
quantized Landau levels. While this scenario may give rise to an integer
quantum Hall effect, the FQHE requires repulsive interactions between the
particles as its crucial ingredient. In the context of FQHE as a solid-state
phenomenon, almost exclusively Coulomb interactions have been considered. For
neutral fermions, however, it is most likely to have dipolar interactions. We
therefore study the following Hamiltonian ($\hbar=1$):
\begin{equation}
\label{H}
\Hat{\mathrm{H}}  = \sum_{j=1}^{N} \frac{1}{2m} \left[  \vec{p}_{j}-\vec{A}(\vec{r}_{j})\right]  ^{2} +\sum_{j<k}^{N} V_{\mathrm{dd}}(\vec{r}_{j},\vec{r}_{k}).
\end{equation}
Here $m$ is the mass of the particles and $V_{\mathrm{dd}}(\vec{r}_{j},\vec
{r}_{k}) = d^{2} / |\vec{r}_{j} -\vec{r}_{k}|^{3}$ the dipolar interaction
between the particles with dipolar moment $d$. The gauge potential $\vec{A}$
is supposed to describe a constant gauge field perpendicular to the
two-dimensional system, and will be specified in Section \ref{III}.

It is known that for practically any repulsive interaction a very good trial
wave function for the ground-state is the so-called Laughlin function
\cite{girvin-qhe}, which can be found by filling the lowest Landau level up to
a filling factor $\nu$ \cite{laughlin}:
\begin{align}
\label{psilaugh}\Psi_{\mathrm{L}} ( \left\{  z_{j} \right\}  )  & =
{\mathcal{N}} \prod_{k<l}^{N}(z_{k}-z_{l})^{1/\nu} \exp\bigg(-\sum_{i}%
^{N}\left\vert z_{i}\right\vert ^{2}/4l_{0}^{2}\bigg),
\end{align}
where $z_{i} \equiv x_i + i y_i$ are the positions of the particles,
$\mathcal{N}$ is a normalization factor, and $l_{0} \equiv
\sqrt{1/B_{0}}$ is the magnetic length corresponding to a magnetic field of
strength $B_{0}$. If $1/\nu$ is odd, the function is fully antisymmetric and
thus describes fermionic systems, while the opposite is true for even $1/\nu$.

The Laughlin ground-state can be considered a homogeneous liquid state. It is
excited by piercing a hole into it. Choosing for simplicity the quasihole to
be at the origin, the excited state is described by
\begin{align}
\label{psiqh}\Psi_{\mathrm{qh}}(\left\{  z_{j}\right\} )  &  =
{\mathcal{N}}' \prod_{j=1}^{N} z_{j} \Psi_{\mathrm{L}},
\end{align}
with ${\mathcal{N}}'$ a normalization factor.

A gap, vaguely defined as the energy difference separating the Laughlin state
from the quasihole state, guarantees the robustness of the FQHE. But as one
may think of different ways of creating a quasihole, it is important to be
precise in the definition. In a finite system, a quasihole can be created in
three different ways \cite{morf-halperin, girvin}:

(i) by reducing the particle number,

(ii) by increasing the area of the system at constant particle number,

(iii) by changing the magnetic field at constant particle number and constant area.

In the electronic system, in addition to the Coulomb repulsion between the
electrons, a positively charged background (originated from ions) is present
as a consequence of the electroneutrality of the system. The corresponding
background potential stabilizes the Coulomb system, but it also adds to the
energy of the quasihole, compensating the losses in the energy of the direct
electron-electron Coulomb interaction due to lowering the number of electrons
or their density. As a result, for any choice of the definition a positive energy gap is obtained \cite{morf-halperin, girvin}.

Contrary to the electron case, dipolar systems have no such background
potential, and the definition of the gap is crucial. As we will explicitly see
in Section \ref{IV}, energy \textquotedblleft gaps\textquotedblright\ defined
according to (i) or (ii) have a negative sign. This would mean that the system
is unstable against the creation of quasiholes. In fact, this, at first view,
counter-intuitive result is not astonishing: Since (i) and (ii) results in
diluting the system and, therefore, in lowering the potential energy, it is
obvious that in these cases we should find an energy gain. Therefore, the only
meaningful definition for the energy gap in a dipolar system is according to
(iii), where we compare the energy of a quasihole with the ground state energy
of the system with the same number of particles occupying the same area -
similar to the definition of the quasihole excitation energy in a normal Fermi system.

\section{Thermodynamic Approximation \label{ii}}

In order to evaluate the gap, we follow an approach developed in Refs.
\cite{girvin-g0,*girvin}. Based on the plasma analogy \cite{laughlin},
relating the physics of the Laughlin state to the one of a classical
one-component plasma, analytic expressions for the correlation functions
$g_{0}(z_{1},z_{2})$ and $g_{\mathrm{qh}}(z_{1},z_{2})$ of the Laughlin state
and the quasihole state can be derived: \begin{widetext}
\vspace*{-0.7cm}
\begin{align}
\label{g0}
g_{0}(z_1, z_2)  &= \frac{\nu^2}{(2\pi)^2}\!\!\Big(1-{\rm
e}^{-\frac{\abs{z_1-z_2}^2}{2}}
-2\sum_{j}^{\rm \scriptscriptstyle odd}\frac{C_j}{4^jj!}\abs{z_1-z_2}^{2j}{\rm
e}^{-\frac{\abs{z_1-z_2}^2}{4}}\!\!\Big),\\
\label{gqh}
g_{\rm qh}(z_1, z_2) &= \frac{\nu^2}{(2\pi)^2}\bigg[\prod_{j=1}^2
\Big(1-{\rm
e}^{-\frac{\abs{z_j}^2}{2}}\Big)
-{\rm e}^{-\frac{\abs{z_1}^2+\abs{z_2}^2}{2}}
\bigg(\Big|{\rm e}^{\frac{z_1z^\star_2}{2}}-1\Big|^2+
2\sum_{j}^{\rm \scriptscriptstyle odd}\frac{C_j}{4^jj!}\sum_{k=0}^\infty
\frac{\abs{F_{j,\,k}(z_1,\,z_2)}^2}{4^kk!}
\bigg)\bigg],\\
\label{F}
F_{j,k}(z_1, z_2)&=\frac{z_1z_2}{2}
\sum_{r=0}^{j}\sum_{s=0}^{k}
{j \choose r}{k \choose s}
\frac{(-1)^r z_1^{r+s} z_2^{j+k-(r+s)}}{\sqrt{(r+s+1)(j+k+1-(r+s))}}.
\end{align}
\end{widetext}
Note that in these expressions, the quasihole state is related to the Laughlin
state by reducing the particle density in the center of a thermodynamically
large system, which according to the classification of the previous section is
a quasihole of type (i).

Given these expressions, we may evaluate the energy difference $\Delta$
between these two states by solving the following integral
\begin{align}
\label{Delta}\Delta=  &  \frac{1}{2} \int\mathrm{d}^{2} z_{1} \int
\mathrm{d}^{2} z_{2} \ V_{\mathrm{dd}}(z_{1},z_{2})\\
&  \left[  g_{\mathrm{qh}}(z_{1}, z_{2}) - g_{0}( z_{1}, z_{2}) \right]
.\nonumber
\end{align}

Fortunately, we are able to derive a relation between the quantity from Eq.
(\ref{Delta}) and the gap related to a quasihole state according to (iii).
First, we notice that $\Delta=E_{\mathrm{qh}}^{(N-\nu)} - E_{0}^{(N)}$, where
these are the energies of $N-\nu$ particles in a quasihole state, and $N$
particles in the Laughlin state. Now we may write $E_{0} = N \epsilon_{0}$,
where $\epsilon_{0}$ is the energy of one particle in the Laughlin state. This
quantity can readily be evaluated by substituting $z_{-} \equiv z_{1} - z_{2}$
in $V_{\mathrm{dd}}(z_{1},z_{2}) = V(z_{-})$ and $z_{-} \equiv z_{1} - z_{2} $
in $g_{0}(z_{1}, z_{2}) = g_{0}(z_{-})$, and integrating:
\begin{align}
\label{epsilon}\epsilon_{0} = \frac{(2 \pi)^{2}}{2 \nu} \int_{-\infty}%
^{\infty} \mathrm{d} z_{-} \ g_{0}( z_{-}) V_{\mathrm{dd}}(z_{-}).
\end{align}

Demanding a constant particle number, we may re-define the gap as
\begin{align}
\label{Delta_N}\Delta_{N} \equiv E_{\mathrm{qh}}^{(N-\nu)} - E_{0}^{(N-\nu)} =
\Delta+ \nu\epsilon_{0}.
\end{align}

This definition describes a quasihole created according to (ii). However, as
argued in the previous section, apart from a fixed particle number, we should
also demand a fixed volume. Therefore we notice that each Landau state
occupies an area $a= \pi l_{0}^{2}$. For the Laughlin state with $N$ particles
to occupy the same area as the quasihole state with $N$ particles and one
quasihole, we thus have to modify the magnetic length $l_{0}^{\prime}$ of the
excited state according to:
\begin{align}
\label{frac}
\frac{l_{0}^{\prime 2}}{l_{0}^{2}} = \frac{N}{N+\nu}.
\end{align}
Now we have to note that the energies in the dipolar system scale with
$l_{0}^{-3}$. Since we wish to compare states at different magnetic fields, we
define the gap at constant particle number and constant volume as:
\begin{align}
\frac{\Delta_{V}}{l_{0}^{3}} = \frac{E_{\mathrm{qh}}^{(N)}}{l_{0}^{\prime 3}} -
\frac{E_{0}^{(N)}}{l_{0}^{3}}.
\end{align}
By noticing that
\begin{align}
\frac{E_{\mathrm{qh}}^{(N)}}{l_{0}^{\prime 3}} = \left(  \Delta\frac{l_{0}^{3}%
}{l_{0}^{\prime 3}} + N \epsilon_{0} \frac{l_{0}^{3}}{l_{0}^{\prime 3}}
\right)  \frac{N}{N-\nu},
\end{align}
and approximating $N/(N-\nu) \approx(N+\nu)/N$ for large $N$, we find with Eq.
(\ref{frac}):
\begin{align}
\label{Delta_V}\Delta_{V} = \Delta+ \frac{5}{2} \nu\epsilon_{0}.
\end{align}
Before we calculate this quantity, we discuss different gauge potentials that
might be realized.

\section{ \label{III} Different Gauge Potentials}

For the discussion of FQHE, it is feasible to choose a symmetric gauge for the
gauge field. A perpendicular, magnetic field of strength $B_{0}$ is then
described by the magnetic vector potential $\vec{A}_{\mathrm{mag}}(\vec{r}%
_{j}) = B_{0}(-y,x)$. In the context of artificial fields, it is possible to
generalize this potential to be an element of $SU(2)$. We choose it to have the form:
\begin{align}
\label{A}\vec{A}(\vec{r}_{j}) = B_{0}(-y,x)+(\alpha\sigma_{y},\beta\sigma
_{x}),
\end{align}
where $\alpha$, and $\beta$ are additional, controllable parameters, and
$\sigma_{x,y}$ are Pauli matrices. Proposals to realize such gauge potentials
have been made both for lattice systems \cite{osterloh} and for trapped gases
coupled to laser fields \cite{fleischhauer}. Eq. (\ref{A}) contains the limit
of a magnetic field, as can be seen by choosing $\alpha=\beta=0$. For finite
$\alpha$ and $\beta$, the potential (\ref{A}) yields a constant non-Abelian
gauge field perpendicular to the system. For the isotropic configuration
$\alpha= \beta$ it has been shown in Ref. \cite{trombettoni} that this
potential splits the Landau levels, which, by assuming a short-range repulsion
between the particles, may give rise to a FQHE with non-Abelian anyons. The
most general configuration with $\alpha\neq\beta$ has been considered for
non-interacting lattice systems in Ref. \cite{anjaPRL}. Then the $SU(2)$ gauge
potential produces an anisotropic space-time. Giving rise to Dirac points in
the bandstructure of the free system, we may measure this anisotropy as the
ratio of the sound velocities in $x$- and $y$-direction: $c_{x}/c_{y} =
|\alpha|/ |\beta|$. In Ref. \cite{anjaPRL} it has been shown that this
situation is best described by introducing a squeezing parameter
\begin{align}
\xi= - \text{tanh}^{-1}\left(  \frac{c_{y}-c_{x}}{c_{y}+c_{x}} \right) ,
\end{align}
and replacing the original variable $z$ by squeezed one $\tilde z$:
\begin{align}
\label{squeeze}\big[ z \equiv x + i y \big] \rightarrow\big[ \tilde z(\xi)
\equiv\cosh\xi\ z - \sinh\xi\ \bar z \big].
\end{align}

We are then able to treat both the Abelian scenario and the non-Abelian
scenario on the same footing: The generalization of the Laughlin wavefunction
and the quasihole wavefunction, Eqs. (\ref{psilaugh}) and (\ref{psiqh}), to
systems with a non-Abelian field are straightforwardly given by making the
replacement Eq. (\ref{squeeze}). Accordingly, the correlation functions
derived for the states Eqs. (\ref{psilaugh}) and (\ref{psiqh}), also hold for
the corresponding squeezed states, if we again make the substitution Eq.
(\ref{squeeze}). The gap, as defined in Eq. (\ref{Delta}) can then be
evaluated by the integral
\begin{align}
\label{Delta2}\Delta(\xi) =  &  \frac{1}{2} \int\mathrm{d}^{2} z_{1}
\int\mathrm{d}^{2} z_{2} \ V_{\mathrm{dd}}(z_{1},z_{2})\\
&  \left[  g_{\mathrm{qh}}(\tilde z_{1}(\xi), \tilde z_{2}(\xi)) -
g_{0}(\tilde z_{1}(\xi), \tilde z_{2} (\xi)) \right] .\nonumber
\end{align}
In the same way, we generalize the ground-state energy defined in Eq.
(\ref{epsilon}) to be a function $\epsilon_{0}(\xi)$ of the squeezing.
Following the derivation as in Section \ref{ii}, we finally arrive at the
equation $\Delta_{V}(\xi)= \Delta(\xi) + \frac{5}{2} \nu\epsilon_{0}(\xi)$.

\section{\label{IV} Results}

While in the Abelian scenario with $\xi=0$ parts of the calculation
can be done analytically, the squeezed scenario is more complicated as it
demands a fully numerical treatment. Since in both cases the steps of the
calculation are the same, we describe in details only the procedure for the
Abelian case.

The main difficulty consists in evaluating the integral Eq. \ref{Delta}. First
we have to specify the coefficients $C_{j}$ in Eqs. (\ref{g0}) and
(\ref{gqh}).  It is shown in Ref. \cite{girvin-g0} that by setting all $C_{j}
= 0$, a system with a completely filled Landau level is described, $\nu=1$.
For this choice of $C_{j}$, the resulting correlation functions are denoted by
$g_{0}^{(1)}$ and $g_{\mathrm{qh}}^{(1)}$. In order to have a FQHE, we need a
fractional filling, $\nu=1/q$, which requires the coefficients $C_{j}$ with $j
\leq q$ to be non-zero. For fermions, the most robust effect is expected for
$\nu=1/3$, where the choice $C_{1}=1$ and $C_{3}=-1/2$ is best suited. We call
the corresponding correlation functions $g_{0}^{(3)}$ and $g_{\mathrm{qh}%
}^{(3)}$, and also define for convenience the differences $\Sigma_{0} \equiv
g_{0}^{(3)} - g_{0}^{(1)}$ and $\Sigma_{\mathrm{qh}} \equiv g_{\mathrm{qh}%
}^{(3)} - g_{\mathrm{qh}}^{(1)}$.

Turning now to the integral Eq. (\ref{Delta}) with $\nu= 1/3$, we note that in
the Abelian limit it reduces to the one considered in Ref. \cite{baranov}. As
our numerical result, however, drastically differs from Ref. \cite{baranov}, a
careful analysis is of order. Therefore we split the integral Eq.
(\ref{Delta}) into two parts, $P_{1} \equiv\int\mathrm{d}z_{1} \int
\mathrm{d}z_{2} \ V_{\mathrm{dd}} (g^{(1)}_{\mathrm{qh}} -g_{0}^{(1)})$, which
is analytically solvable, and $P_{2} \equiv\int\mathrm{d}z_{1} \int
\mathrm{d}z_{2} \ V_{\mathrm{dd}} (\Sigma_{\mathrm{qh}} -\Sigma_{0}^{(1)})$,
which we treat numerically.

For the analytic part we find $P_{1} = -\sqrt{2\pi}/\nu^{2} \frac{d^{2}}%
{l_{0}^{3}}$. Note that for $\nu=1$, this negative number would be the full,
completely analytic result for the energy difference $\Delta$ defined in Eq.
(\ref{Delta}). This clearly shows what we have anticipated in Section
\ref{II}, namely that this definition is not the appropriate one for the
energy gap in a dipolar system.

Before we evaluate $P_{2}$ numerically, we examine the asymptotic behavior of
the integrand. As the divergence in the interaction term for $z_{1}
\rightarrow z_{2}$ is compensated by the vanishing of the correlations, this
limit can easily be handled by a regularization of the integral. The limit of
$z_{+} \equiv z_{1} + z_{2} \rightarrow\infty$, however, turns out to be
problematic: For finite particle distance, $|z_{1}-z_{2}|<\infty$, this
contribution is not suppressed by the interaction, and the convergence of the
integral Eq. (\ref{Delta}) requires that $\Sigma_{0}$ and $\Sigma
_{\mathrm{qh}}$ have the same asymptotic behavior. However, the completely
different structure of both functions obscure the latter. Contrariwise, we
should note
that if we truncate the infinite sum in $\Sigma_{\mathrm{qh}}$, this
expression gets exponentially damped for large center-of-mass coordinates,
while $\Sigma_{0}$ depends only on the relative coordinates, yielding
$\Sigma_{0} - \Sigma_{\mathrm{qh}} \neq0$ for $|z_{+}| \rightarrow\infty$.

To circumvent this problem, we bring $\Sigma_{0}$ to a form similar as
$\Sigma_{\mathrm{qh}}$, which is possible by factoring out a damping
$\mathrm{exp}[-(|z_{1}|^{2}+|z_{2}|^{2})/2]$ and Taylor expanding the
remaining exponential $\mathrm{exp}[|z_{+}|^{2}]$. We are then able to write
\begin{align}
\label{F0}\Sigma_{0}(z_{1},z_{2})  & = \mathrm{e}^{-\frac{\left|  z_{1}
\right| ^{2}+\left|  z_{2} \right| ^{2}}{2}} \sum_{j} \frac{-2 C_{j}}{4^{j}
j!} \sum_{k=0}^{\infty} \frac{|F_{j,k}^{(0)}(z_{1},z_{2})|^{2}}{4^{k} k!},\\
F_{j,k}^{(0)}(z_{1},z_{2})  & = \sum_{r=0}^{k} \sum_{s=0}^{k} {\binom{j }{r}%
}{\binom{k }{s}} (-1)^{j-r} z_{1}^{r+s} z_{2}^{j+k-(r+s)}.
\end{align}
As now each term in both $\Sigma_{0}$ and $\Sigma_{\mathrm{qh}}$ is damped by
a factor $\mathrm{exp}[-(|z_{1}|^{2}+|z_{2}|^{2})/2]$, they all vanish in the
limit $|z_{+}| \rightarrow\infty$, and we may truncate the infinite sums at a
sufficiently large value of $k$. Note that due to the different orders in
$z_{1}$ and $z_{2}$ of $F_{j,k}$ in Eq. (\ref{F}) and $F_{j,k}^{(0)}$ in Eq.
(\ref{F0}), the sum in $\Sigma_{0}$ should contain two more terms than the sum
in $\Sigma_{\mathrm{qh}}$ for a quick convergence.

\begin{figure}[ptb]
\includegraphics[width=0.4\textwidth]{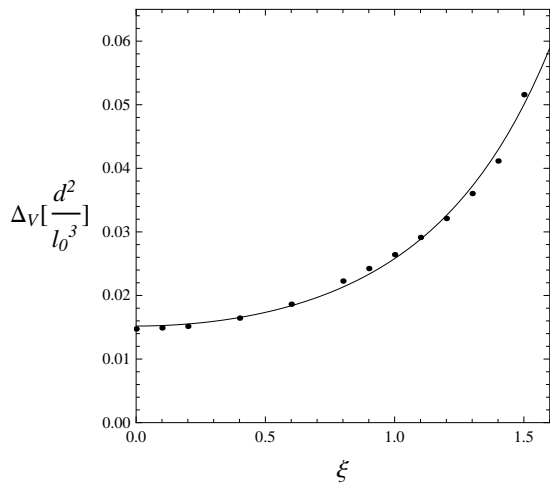}\caption{The gap $\Delta_{V}$ at
constant volume and constant particle number as a function of the squeezing
parameter $\xi$: The dots are obtained by a numerical evaluation of Eq.
(\ref{Delta}) for different $\xi$, a fit of this data yields the solid line.}%
\label{gap}%
\end{figure}

Now we are able to perform the numerical integration. The error due to the
truncation still is $5 \%$ for 10 terms, but can be minimized by a finite-size
analysis of our results. We then find $P_{2} = (0.1875 \pm0.0010) \frac{d^{2}%
}{l_{0}^{3}}$, where the numerical error has been approximated by the
deviation from the smooth fit in Fig. \ref{gap}. With this, we find $\Delta=
0.5 (P_{1}+P_{2}) = - (0.0455 \pm0.0010) \frac{d^{2}}{l_{0}^{3}}$. As we have
argued in Section \ref{II}, this negative value is due to the reduced density
of the system.

We continue with calculating the gap as defined in Eq. (\ref{Delta_N}).
Therefore we have to evaluate the integral Eq. (\ref{epsilon}), which in the
isotropic case, $\xi=0$, reads $\epsilon_{0} = \frac{\sqrt{\pi}}{2\nu} \left(
\frac{\sqrt{2}}{2} - \frac{15}{32}\right)  \frac{d^{2}}{l_{0}^{3}}$, from
which we find that also $\Delta_{N} <0$. The negative values for $\Delta_{N}$
can be understood by noticing that as the particle taken away at the origin
now has been added at the edge of the system increasing its volume, so
$\Delta_{N}$ corresponds to the energy of a quasihole according to (ii). As
long as such a process is possible, the system is unstable as it tries to
reduce its density by diluting.

Finally, we turn to definition Eq. (\ref{Delta_V}). Only in this case, we
obtain a positive gap, $\Delta_{V} = (0.0132 \pm0.0020) \frac{d^{2}}{l_{0}%
^{3}}$, which however is much smaller than the number found in Ref.
\cite{baranov}, $(0.9271 \pm0.019) \frac{d^{2}}{l_{0}^{3}}$, but compares well
with the gap obtained via exact diagonalization of a small dipolar system in
Ref. \cite{osterloh-exact}, where the discrepancy to Ref. \cite{baranov} has
been attributed to the different system size.

Repeating this procedure for finite squeezing $\xi$, we have to evaluate
$\Delta(\xi)$ as defined in Eq. (\ref{Delta2}), but now the whole integral has
to be solved numerically. Again we find negative values for $\Delta(\xi)$ and
$\Delta_{N}(\xi)$, which even decrease with larger $\xi$. However, as also the
ground-state energy $\epsilon_{0}(\xi)$ increases with $\xi$, the gap
$\Delta_{V}(\xi)$ at constant particle number and constant volume finally has
a positive balance for all $\xi$. As shown in Fig. \ref{gap}, it increases
with $\xi$, and a convenient fit to the numerical data is found to be:
\begin{align}
\Delta_{V} (\xi) = \Delta_{V}(0) \exp\left( \alpha\ \xi^{2} \right) .
\end{align}
We obtain $\alpha= 0.529$ and $\Delta_{V}(0)=0.0152 \ d^{2}/l_{0}^{3}$.

To understand this behavior, we note that the squeezing allows the particles to
get closer in one direction, while the particle distance is increased in the
other direction. Due to $1/r^3$ behavior of the dipole-dipole interaction, the
interaction energy is much more sensitive to changes of the density distribution
at short distances rather than at large ones. Thus, compressing in one direction
and stretching in another one increases the interaction energy. As a consequence
of Eq. (\ref{Delta_V}), this gives rise to a bigger energy gap.

\section{Conclusion}

Summarizing this work, we have shown that dramatically differing from the
predictions in Ref. \cite{baranov}, only a small energy gap separates the
Laughlin state from quasihole excitations in systems of dipolar quantum gases
with artificial magnetic fields. However, by considering scenarios where a
non-Abelian gauge field introduces an anisotropy into the system, an
exponential increase of the gap can be achieved, and may allow for robust
fractional quantum Hall states.

\section*{Acknowledgements}

We acknowledge financial support from the ERC Grant QUAGATUA, the ESF
Programme EUROCORES entitled 'Cold Quantum Matter', the Spanish project MINCIN
FIS 2008-00784, the EU projects AQUTE and NAMEQUAM, the Humboldt Foundation,
and the Hamburg Award for Theoretical Physics.

\end{document}